\documentclass{aa}
\usepackage{graphics}
\begin{document}
   \thesaurus{03     %
              (11.03.1;  
               11.05.2;  
               11.09.3;  
               11.19.2)} 

\title{Role of disk galaxies in the chemical enrichment of the
intracluster medium}

\author{D.S.Wiebe \and B.M.Shustov \and A.V.Tutukov}

\offprints{D.Wiebe (dwiebe@inasan.rssi.ru)}

\institute{Institute of Astronomy of the Russian Academy of Sciences,
Pyatnitskaya str. 48, 109017, Moscow, Russia}

\date{Received September 15, 1996; accepted March 16, 1997}

\titlerunning{Chemical enrichment of the intracluster medium}
\authorrunning{Wiebe et al.}

\maketitle

 \begin{abstract}
Elliptical galaxies are often assumed to be the primary source of heavy
elements in the intracluster me\-di\-um (ICM), with the contribution of
other morphological types being negligible. In this paper we argue
that a role of spiral galaxies in the chemical evolution of the ICM is
also important. This statement rests upon our recent calculations of
the heavy element loss from a disk galaxy through the hot steady-state
galactic wind and dust grains expulsion by stellar radiation pressure.
This model reproduces main properties of our Galaxy and, being applied
to galaxies of various masses, explains the observed correlation
between spiral galaxy mass (luminosity) and metallicity. In our model
this correlation develops as a result of the mass dependence of both
loss mechanisms, in the sense that less massive galaxies lose metals
more efficiently. We show that a typical disk galaxy is nearly as
effective in enriching the ICM as an elliptical galaxy of the same
mass.

Having estimated the oxygen and iron loss from a single galaxy, we
integrate them over the galactic mass spectrum. We show that the
`effective' loss (per unit luminosity) from spiral galaxies is
comparable to the loss from ellipticals. The dominant role of
early-type galaxies in rich clusters is caused by that they outnumber
spirals. We present some arguments to this point, based on recent
determinations of the ICM abundances, emphasizing the fact that the
ratio of total iron mass to cluster luminosity does not depend on the
fraction of cluster spirals in a wide range of the latter, contrary to
what one might expect if spirals do not contribute into the ICM
$Z$-abundance.

 \end{abstract}

\section{Introduction}

Clusters of galaxies are believed to be the only closed chemically
evolving systems in the Universe, in the sense that they do
not lose matter
into the intercluster space. Inside them, however, violent
processes of matter and energy exchange between the galaxies and the
intracluster medium should occur. This conclusion comes from the
fact that heavy elements, which are nearly equally distributed between
galaxies and the ICM, can only be produced within galaxies in the course
of `normal' stellar evolution (e.g. Renzini \cite{R97}).

Since in recent years the number of the ICM abundance determinations
and their accuracy have increased significantly, the interest to the
field of the chemical evolution of galaxy clusters has raised, too.
Several mechanisms have been suggested to explain the metal
transport from the galaxies to the ICM. Among them are enriched gas
ejection during the merging of protogalactic fragments (protogalaxies)
in the early Universe (Gnedin \cite{Gnedin}), ram-pressure stripping
of enriched gas both from spiral and elliptical galaxies (Gunn \& Gott
\cite{Gunn}; Himmes \& Biermann \cite{HB80}; Fukumoto \& Ikeuchi
\cite{FI96}), and the hot galactic wind from ellipticals (Matteucci \&
Vettolani \cite{MV88}; David et al. \cite{DFJ91}; Matteucci \& Gibson
\cite{MG95}; Renzini \cite{R97}; Gibson \& Matteucci \cite{GM97}; and
others).

From these, the last mechanism draws more and more attention as an
ultimate explanation of the ICM $Z$-abun\-dance. Moreover, as Arnaud et
al. (\cite{Aetal92}) pointed out, it was the hypothesis about the
galactic wind from ellipticals, which led to prediction of heavy
element existence in the ICM (Larson \cite{L74}; Larson \& Dinerstein
\cite{LD75}). This hypothesis arises from the assumption that
ellipticals began their history with the strong starburst followed by
multiple supernova events. These explosions after a very short time
expelled all the interstellar matter into the ICM, having provided it
with heavy elements and thermal energy. At the same time, hot galactic
wind is thought to be responsible for establishing the correlation of
luminosity (mass) and metallicity, observed in ellipticals.

As several theoretical studies of this process have shown, mass loss
from ellipticals is effective enough to explain $Z$-abundance in the
ICM, though this mechanism, according to some of these studies,
requires the IMF of high $z$ galaxies to be richer in massive stars
than today (Matteucci \& Vettolani \cite{MV88}; David et al.
\cite{DFJ91}; Matteucci \& Gibson \cite{MG95}; Okazaki et al.
\cite{oetal93}). Futhermore, Arnaud et al. (\cite{Aetal92}) found
observational confirmation of the ellipticals being the dominant
source of intracluster metals in the apparent increase of the iron
mass in galaxy clusters with the growth of the combined E+S0
luminosity. From this they concluded that the role of spiral galaxies
in the ICM enrichment is negligible.

This possibly explains why spiral galaxies lack attention as a
potential source of the intracluster metals, though, as Fukumoto \&
Ikeuchi (1996) argued, a single spiral galaxy could be as effective
in the ICM enrichment as an elliptical one, and correlation of iron
mass with the combined luminosity of ellipticals and lenticulars is
caused by that early-type galaxies outnumber spirals. Furthermore,
recently Kauffmann \& Charlot (\cite{kc98}) suggested that metals in
the ICM had been ejected from the primordial population of spirals
that have later merged to form present-day early-type galaxies. In
their model, spiral galaxies are not the dominant, but the only source
of heavy elements in hot intracluster gas.

Recently we also argued that disk galaxies suffer significant $Z$-loss
(Shustov et al. \cite{weare97}; hereafter Paper I). We considered two
processes leading to heavy element expulsion from disk galaxies:
hot galactic wind and dust expelling by the stellar radiation
pressure. It was shown that these processes are rather effective and
may result in the luminosity-metallicity correlation observed in disk
galaxies. In this paper we extend these calculations to the
integration of heavy-element production in disk galaxies of various
masses over the galactic mass function, thus assessing $Z$-loss from
all disk galaxies in a typical cluster.

In Section~2 we briefly describe the chemical evolution model used
for computation of heavy element loss from a disk galaxy. In
Section~3 results of integration of oxygen and iron losses over the
galactic mass function are presented and compared with the analogous
data for elliptical galaxies. They are discussed in Section~4. A
conclusion follows that a disk galaxy enriches the ICM as effective as
an elliptical one of the same mass.

\section{Chemical evolution model}

Model of galactic evolution used here is outlined in Paper~I and Wiebe
et al. (\cite{weare98}). Its primary feature is that in addition to
the conventional mass conservation equations for a total gas mass and
for a particular element (Tinsley \cite{T80}) we solve the energy
equation, based on a balance between the turbulent energy input from
supernova explosions and its dissipation in cloud-cloud collisions. In
Paper I we have successfully used this model to show that disk
galaxies suffer significant loss of heavy elements, while in Wiebe at
al. (\cite{weare98}) we showed that this model is good in reproducing
main characteristics of our Galaxy.

As in Wiebe et al. (\cite{weare98}), here we model the evolution of
oxygen and iron. For the oxygen production we interpolate
metallicity-dependent yields of Maeder (\cite{maeder}). Iron is
assumed to originate both in core-collapse and Ia supernovae. Its
production is taken into account according to Thielemann, Nomoto \&
Hashimoto (\cite{iron1}; SN~Ib+II) and Tsujimoto et al. (\cite{iron1};
SN~Ia). Each Ia~Supernova is assumed to produce 0.6$M_\odot$ of iron.
Time delay for these SNe is taken to be equal to $3\times10^8$~years
(Tutukov \& Yungelson \cite{ty94}). We use Salpeter-like initial mass
function with limits 0.1 and 100$M_\odot$.

\begin{figure*}
\resizebox{0.8\hsize}{!}{\includegraphics{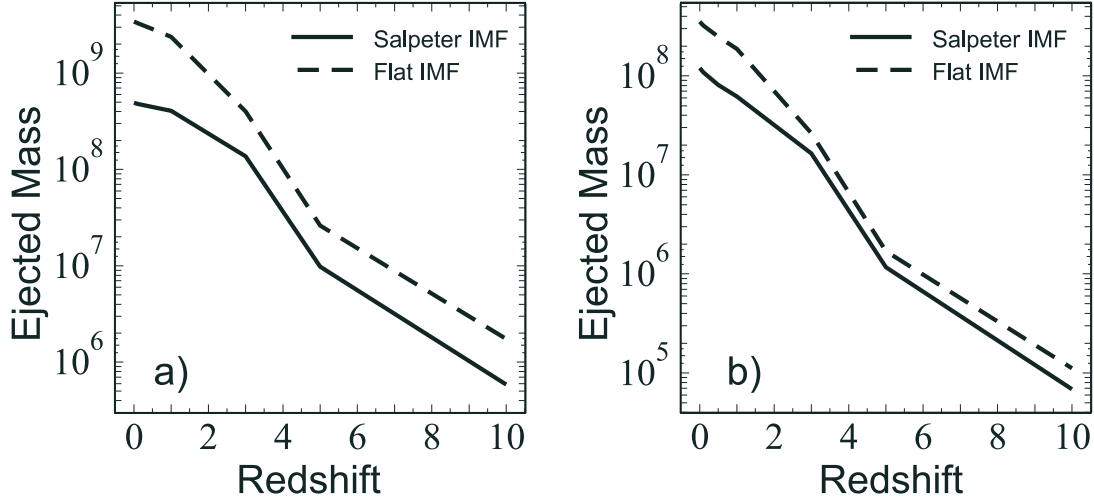}}
\caption{Redshift (time) dependence of oxygen (a) and iron (b) masses
ejected from a galaxy with $M_{\rm G}=5\cdot10^{11}M_\odot$.}
\label{OFe1}
\end{figure*}

\begin{figure*}
\resizebox{0.8\hsize}{!}{\includegraphics{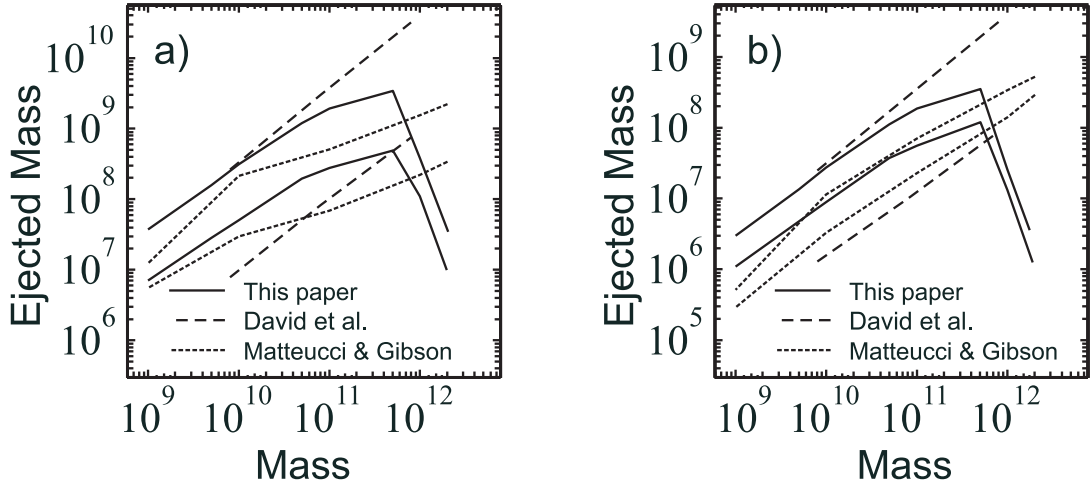}}
\caption{Dependence of theoretical oxygen (a) and iron (b) ejected mass
on the mass of a galaxy for spirals (this paper) and ellipticals (previous
investigations).}
\label{OFe2}
\end{figure*}

Within the context of the present paper it is important to remind how
we take into account heavy element loss from disk galaxies. We
consider two processes ejecting $Z$ into the ICM. They are hot
galactic wind and dust expelling by the stellar radiation pressure.
Heavy element loss rates due to both processes depend a on galaxy
mass, in the sense that low-mass galaxies lose their metals more
effectively.

Hot galactic wind forms from merged galactic fountains produced by
multiple supernova explosions in OB-associations (Mac Low et al.
\cite{mcl}; Norman \& Ikeuchi \cite{ni89}; Igumentschev, Shustov, \&
Tutukov \cite{I90}). Only a fraction $f_{\rm esc}$ of this wind is
lost into the circumgalactic space. Combining observational and
theoretical results, we derived in Paper~I linear dependence of
$f_{\rm esc}$ on the galaxy mass. This dependence is based on
two theoretical and several observational points. According to
Igumentschev et al. (\cite{I90}), $f_{\rm esc}\ll1$ for galaxies with
masses greater than $10^{12}M_\odot$. De Young \& Gallagher
(\cite{dyg1990}) showed that in a galaxy of $1.4\times10^9M_\odot$
$f_{\rm esc}\le0.6$. In Paper~I we supplied these theoretical limits
with several intermediate observational points, assuming that the
surface filling factor of supershells produced by the multiple
explosions can be a natural estimate for $f_{\rm esc}$. Resultant
dependence of $f_{\rm esc}$ on the logarithm of the galaxy mass is
well fitted by a straight line (see Paper~I for details).

The second process leading to $Z$-loss from a disk galaxy is dust
expelling by the stellar radiation pressure. Existence of extended
dust halo around disk galaxies is demonstrated in a number of
observations (Zaritski \cite{Zar}; Howk \& Savage \cite{HS97}). Using
numerical model elaborated by Shustov \& Wiebe (\cite{dust}), we
estimated dust loss dependence on a galaxy mass. It turns out that
effectiveness of dust expelling decreases with galactic mass, again
approaching nearly zero for galaxies with masses $\sim10^{12}M_\odot$.

A new important feature is included in the present version of our
model. In its previous version, described in Wiebe
et al. (\cite{weare98}), the galaxy forms at once that leads to high
star formation rate in the early galaxy life. To make our model more
realistic, in its present version we change the initial conditions by
adding the initial accretion phase.

In a collapsing (protogalactic) cloud, the rate of matter accretion
onto the cloud core depends only on the cloud initial temperature
(e.g. Massevitch \& Tutukov \cite{fez}). We assume in our standard
model that the accretion rate is 100$M_\odot$ yr$^{-1}$. That
corresponds to the accretion time $2\times10^9$ years. During this
time mass of the Galaxy grows with constant rate; when it equals to
$M_{\rm G}$, the accretion stops. This makes the starburst lower and
smoother.

\section{Results}

We perform calculations as follows. Using our Galaxy as a standard,
we find proper values of free parameters to reproduce its main
properties, including dependence of metallicity on the height
above the galactic disk and [O/Fe]--[Fe/H] correlation. After
that, adopted values of parameters are kept fixed and used further
to model galaxies of different masses.

We investigate the evolution of oxygen and iron ejection from
galaxies with masses between $10^{10}$ and $5\cdot10^{12}M_\odot$.
Less massive galaxies also eject heavy elements into the ICM,
however, as we noted in Wiebe et al. (\cite{weare98}), our model
probably does not work for them, mainly because the assumption of the
self-regulated star formation is not valid in small systems.
In order to compare our results with existing estimates of heavy
element ejection  from elliptical galaxies we compute two
models: Salpeter IMF ($x=1.35$) and flat IMF ($x=0.95$).

Shown in Fig.~\ref{OFe1} is a time dependence of ejected oxygen and
iron mass from the galaxy with $M_{\rm G} =5\cdot10^{11}M_\odot$ for
the two different cases (throughout the paper the standard cosmological
model with $h=0.5$ is adopted). Introducing in the model the flat IMF
with higher proportion of massive star increases both oxygen and iron
production nearly by an order of magnitude.

\begin{figure*}
\resizebox{0.8\hsize}{!}{\includegraphics{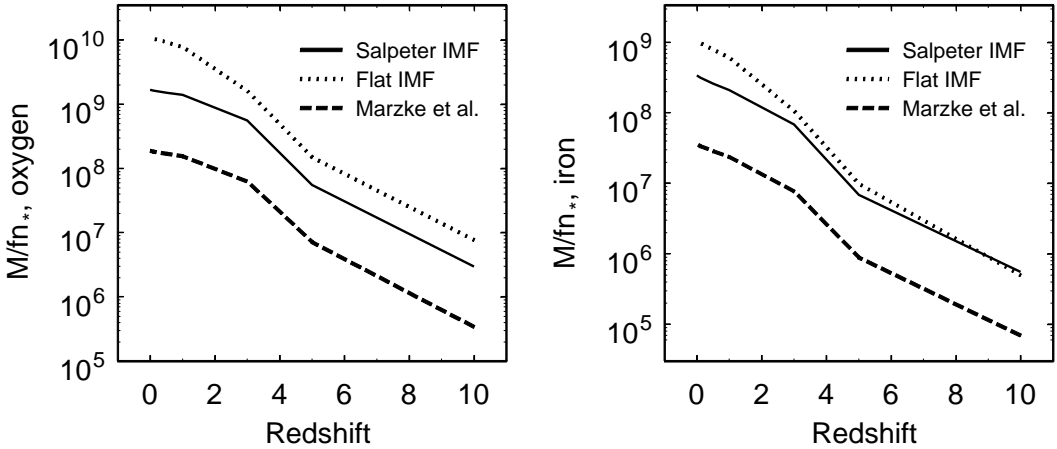}}
\caption{Integrated oxygen (a) and iron (b) ejected mass vs.
redshift for different choices of stellar and galactic mass function.}
\label{OFe3}
\end{figure*}

In Fig.~\ref{OFe2}a our results for oxygen loss (integrated over time
to $t=13\cdot10^{9}$ years) are presented in comparison with
predictions of David et al. (\cite{DFJ91}) and Matteucci \& Gibson
(\cite{MG95}). Fig.~\ref{OFe2}b depicts results for iron ejection
according to the same papers. In all cases upper curve corresponds to
the flat IMF, and lower curve corresponds to the `ordinary' IMF
($x=1.35$ for this paper and Matteucci \& Gibson; $x=2.0$ for David et
al.).

As it is demonstrated by Fig.~\ref{OFe2},
disk galaxies with masses less than $5\cdot10^{11}M_\odot$
are as effective in enriching the ICM with O and Fe as ellipticals.
Specific feature of our model is steep decrease in the heavy element
ejection rate for more massive galaxies. This is caused by the adopted
form of dependence of the ejection efficiency on the galactic mass, as
explained above. The similar decrease of the heavy element ejection
rate for massive disk galaxies was argued also by Kauffmann \& Charlot
(\cite{kc98}). In their model, galaxies with circular velocities greater
than 250 km s$^{-1}$ (i.e., slightly more massive than the Galaxy)
eject only 20 per cent of the intracluster metals.

To assess the overall production of O and Fe in disk galaxies one has to
integrate mass of the element ejected from a single galaxy over the galactic
mass function. It is now widely assumed that the present galactic luminosity
function (LF) can be described by Schechter (\cite{S76}) law with reasonable
accuracy. We take Schechter function in its
original form, assuming that the spiral luminosity function is proportional
to it (universality of LF for all morphological types was recently confirmed
by Andreon \cite{A98} and Marzke et al. \cite{Me98})
\begin{equation}
N=fn_*\int\limits_{l_1}^{l_2}\left({L\over L_*}\right)^{\alpha}
\exp\left[-\left({L\over L_*}\right)\right]
\mbox{d}\left({L\over L_*}\right),
\label{schechter}
\end{equation}
with limits $l_1$ and $l_2$ being the minimum and maximum
luminosities, expressed in units of $L_*$. Here $f$ is a number
fraction of spirals and $n_*$ is cluster richness, as defined by
Schechter (\cite{S76}). One of the quantities computed in our model is
final galaxy luminosity. Comparing it with the galaxy mass, we find
that in our model the present-day mass-luminosity relation fits the
power law
\[
L=1.3M^{0.9}.
\]
Substituting it in (\ref{schechter}), we obtain the present-day mass
function
\[
N=0.9fn_*\int\limits_{m_1}^{m_2}\left({M\over M_*}\right)^{0.9\alpha-0.1}\\
\exp\left[-\left({M\over M_*}\right)^{0.9}\right]
\mbox{d}\left({M\over M_*}\right),\\
\]
where $m_i$ are mass limits mentioned above and $M_*=(L_*/1.3)^{1/0.9}$. Now
we may write the formula for the mass of this
element ejected by all galaxies in the cluster
\begin{eqnarray}
\nonumber M_{\rm Z}^{\rm tot}=0.9fn_*
\int\limits_{m_1}^{m_2}\left({M\over
M_*}\right)^{0.9\alpha-0.1}\times\\
\times\exp\left[-\left({M\over M_*}\right)^{0.9}\right]M_{\rm Z}
\mbox{d}\left({M\over M_*}\right),
\label{eje}
\end{eqnarray}
where $M_{\rm Z}$ is the mass of heavy elements ejected
by the galaxy of the mass $M$.
Using this function we can compute the evolution
of the intergalactic oxygen and iron produced in disk galaxies.

Results of integration are presented in Fig.~\ref{OFe3} where
oxygen and iron yields are shown as functions of redshift. They are
computed with $\alpha=-1.3$ and $L_*$ corresponding to absolute
magnitude $M_{\rm B}=-22.5$ (rich cluster case
from Matteucci \& Gibson \cite{MG95}). As it was the case for a single
galaxy, again the amount of the ejected element is increased by an
order of magnitude for the flat IMF. Note that to convert numbers on
$y$-axis into the total mass of the element ejected in the ICM by all
disk galaxies in the cluster, one should multiply them by $fn_*$.

To compare our results with the predictions about O and Fe ejection
from elliptical galaxies presented by Matteucci \& Gibson
(\cite{MG95}) we computed two models corresponding to their
cases of a rich and poor cluster. Parameters of these models and their
results are listed in Table~\ref{tabref}.
It is obvious that elliptical
galaxies are the dominant source of heavy metals in rich clusters,
while in poor clusters, rich in spirals, the latter ones catch up with
ellipticals in iron production and exceed them in oxygen production.
We shall return to this point in the next section.

To check dependence of our results on the adopted parameters of the LF
we computed oxygen and iron yields with $\alpha$ and $L_*$ from Marzke
et al. (\cite{Me98}). Evolution of O and Fe ejected mass for this LF
is also shown in Fig.~\ref{OFe3}. In this case
production of both iron and oxygen is decreased by an order of
magnitude. However, predicted mass of O and Fe ejected into the ICM by
ellipticals would decrease, too, so this cannot change results of
comparison for the two morphological types.

\begin{table*}
\caption{Predicted mass of elements ejected into the ICM from all
cluster E+S0 galaxies and spirals.}
\label{tabref}
\begin{tabular}{lrrrrcccc}
\hline
Cluster & $\alpha$ & $M_{\rm B}^*$ & $n_*$ & $f$ & \multicolumn{2}{c}{E+S0} & \multicolumn{2}{c}{Spirals} \\
            &                &                          &           &       &   O & Fe & O & Fe \\
\hline
Rich & $-$1.3 & $-$22.5 & 115 & 0.2 & 1.10(11) & 4.04(10) & 3.9(10) &7.9(9) \\
Poor & $-$1.3 & $-$22.0 & 20 & 0.7 &  3.77(9) & 1.19(10) & 1.9(10) & 3.8(9) \\
\hline
\end{tabular}
\end{table*}

\section{Discussion}

Hypothesis on a hot galactic wind associated with the burst of star
formation in elliptical galaxies, being satisfactory in explaining
temperature and abundance of the ICM, contradicts to an apparent lack
of bright ellipticals in the young Universe, found in various deep
field surveys (Kauffmann, Charlot, \& White \cite{KCW96}; Zepf
\cite{Zepf}). A number of effects has been proposed to explain this
fact, intragalactic and intergalactic extinction being among them.
Another explanation was suggested by Kauffmann \& Charlot
(\cite{kc98}). They argued that ellipticals did not form in a single
burst of star formation but formed later during merging of disk
galaxies. That is why we do not see their bright early stages.

\begin{figure}
\resizebox{0.8\hsize}{!}{\includegraphics{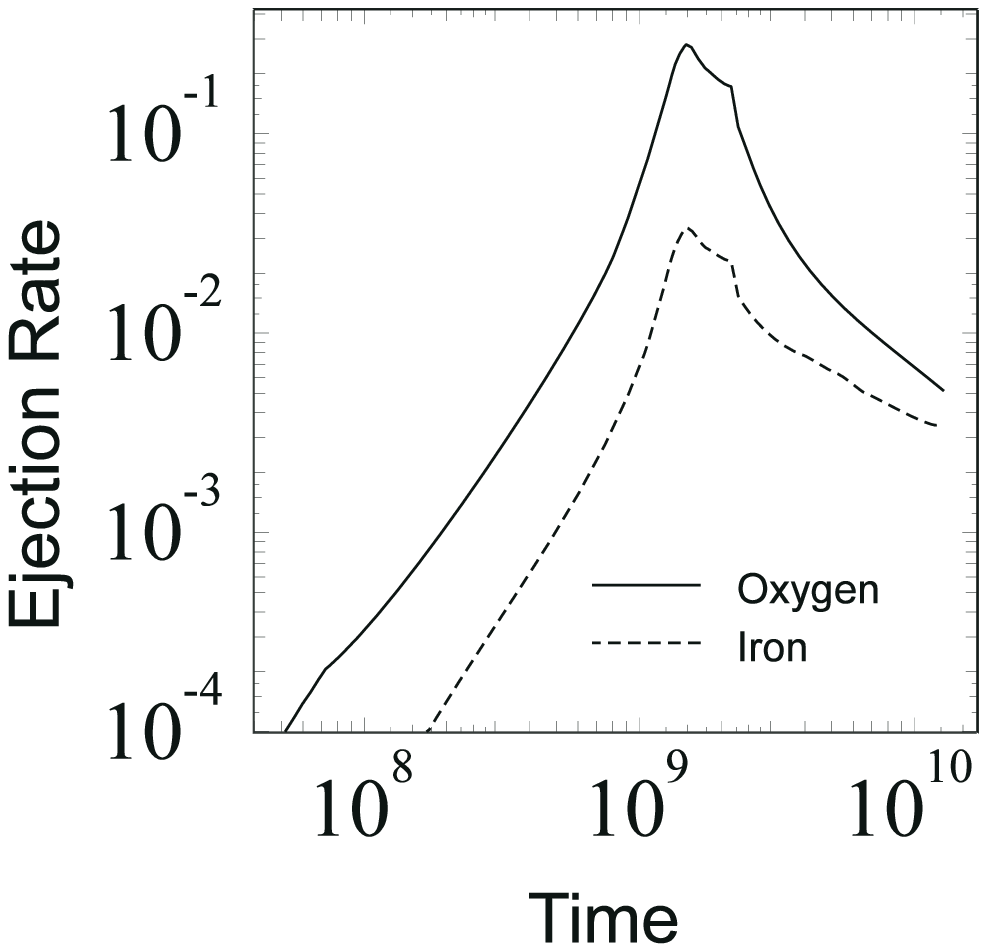}}
\caption{Rate of oxygen (a) and iron (b) ejection for our Galaxy as a
function of time.}
\label{rate}
\end{figure}

In this paper we do not intend to go into details of the elliptical
galaxies formation. Our goal is to investigate the ICM-enriching role
of a single non-interacting disk galaxy. However, our model has to
deal with the problem of the initial starburst, too. Shown in
Fig.~\ref{rate} is the evolution of the rate of heavy element
expulsion for our Galaxy (considered as a standard). It is obvious
that, though the rate is high enough even at the present epoch, the
bulk of $Z$-loss occurs at early stages of galaxy's life as a result
of intense star formation and related high supernova rate. Thus, this
initial starburst is an important factor determining the effectiveness
of heavy element loss from disk galaxies. As we pointed out in Wiebe
et al. (\cite{weare98}), it is impossible to avoid the initial
starburst within the framework of our model without breaking
consistency between the current theoretical parameters of disk
galaxies and observational data. Similar phase of the high star
formation rate (though not so intense as in our model) is obtained,
e.g., in hydrodynamical modelling of our Galaxy by Raiteri et al.
(\cite{rvn96}) and Samland et al. (\cite{sht97}).

There is a growing body of evidence that the star formation bursts did
occur in most galaxies but is hidden from optical surveys by the dust
absorption (see e.g. Blain et al. \cite{blain}). In Wiebe et al.
(\cite{weare98}) we argued that at the time of the starburst a typical
galaxy is able to accumulate enough dust to screen out stellar
radiation. In the most favorable conditions absorption can amount to
several magnitudes.

\begin{figure*}
\resizebox{0.8\hsize}{!}{\includegraphics{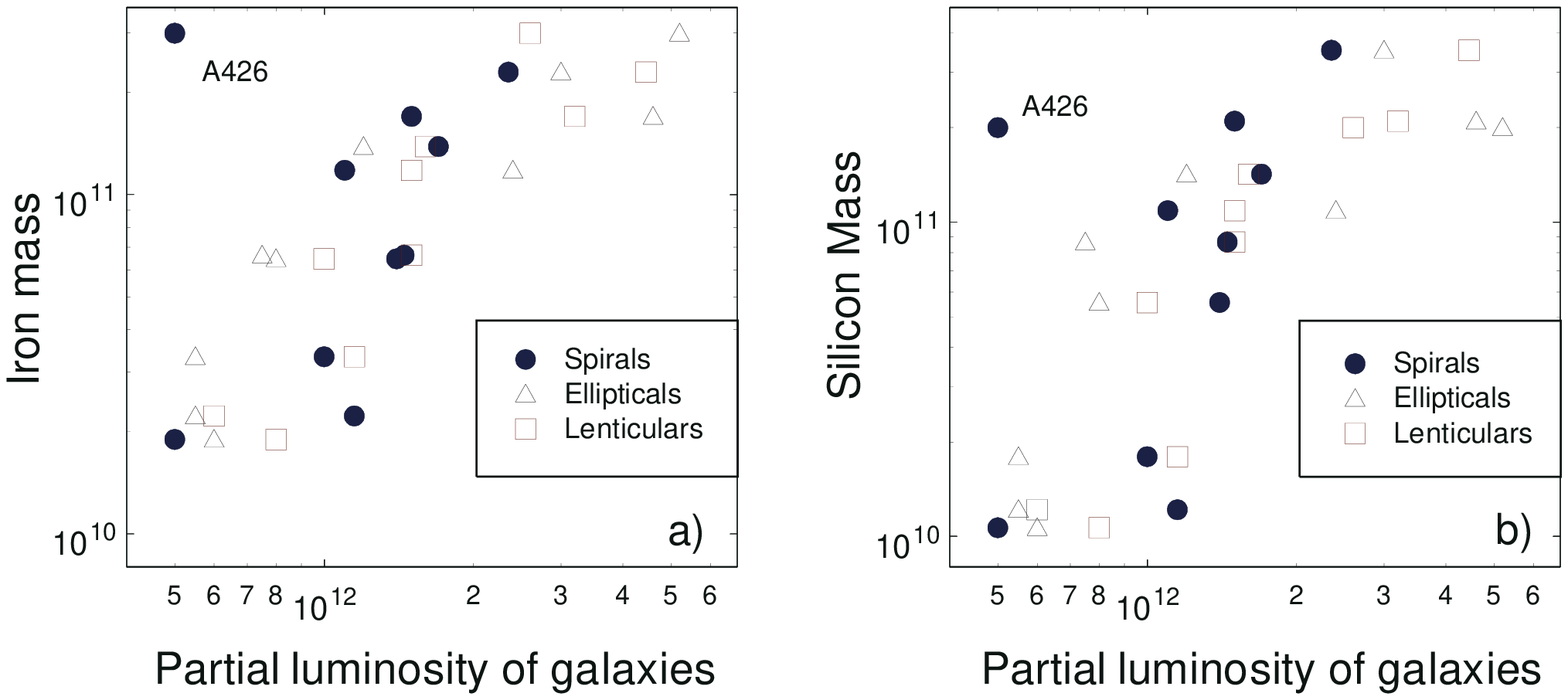}}
\caption{Iron (a) and silicon (b) masses in the galaxy clusters vs.
partial luminosity of galaxies of various morphological types.}
\label{imlr}
\end{figure*}

Another factor, strongly affecting our results, is the used method of
accounting for the dependence of $Z$-loss efficiency on the galactic
mass. The 2D hydrodynamical models of a blow-out of the galactic disk
(Igumentschev et al. \cite{I90}), being a useful instrument for
estimate, cannot be considered as a full treatment of gas dynamics of
heavy element loss from the Galaxy. Note, however, that in dense
gaseous disks, where all processes governing the self-regulated star
formation (shock waves, ionizing radiation, etc.) have relatively
short distance scales, the star formation efficiency is determined by
the local gas parameters and does not depend on global characteristics
of a galaxy. This means that the final gas metallicity within the
framework of our model does not depend on the mass of a galaxy. This
fact is well illustrated by Table~\ref{difmass} where final
metallicities are given with and without taking into account
mass-dependent $Z$-loss. Thus, we have no way of reproducing the
observed mass-metallicity correlation other than mass-dependent
galactic wind and dust grain expelling. Keeping in mind that we are
able to obtain the above correlation in our model (probably,
underestimating $Z$-loss from low-mass galaxies), we may say that this
model is reasonably accurate.

Apparently, we have a contradiction with the results of Arnaud et al.
(\cite{Aetal92}), who argued for the hot wind from ellipticals as the
only significant mechanism of ICM enrichment with heavy elements.
According to our data, both elliptical and disk galaxies give
comparable contributions into the chemical evolution of the ICM.
One of the basic premises of Arnaud et al. (\cite{Aetal92})
paper was the correlation on the intracluster gas mass -- luminosity
diagram for ellipticals and the scatter on the same diagram for
spirals. One may conclude from this fact that ellipticals, in general,
play the dominant role in the intracluster medium evolution, while the
spirals do not. However, as the theoretical modelling of elliptical
galaxies evolution show, all the gas ejected from the cluster
ellipticals (in a typical cluster) cannot account for its amount in
the ICM. So, the bulk of the intracluster gas should have primordial
origin (see Okazaki et al. \cite{oetal93} for discussion), and the
lack of correlation between the intracluster gas mass and spiral
luminosity, being of great interest on its own, has only a little
bearing on the problem investigated.

In 1992 Arnaud et al. had only six clusters with known metallicities
\emph{and} morphological populations. Recently, new data about the ICM
chemical composition have become available (Fukazawa et al.
\cite{F98}). Here we combine them with morphological data compiled by
Arnaud et al. (\cite{Aetal92}) to understand if our results disagree
with the current observations.

\begin{table}
\caption{Present-day oxygen abundance in open and closed models.}
\label{difmass}
\begin{tabular}{ccccccc}
\hline
Mass, &\multicolumn{4}{c}{Oxygen abundance [O/H]}\\
$\log(M_{\rm G}/M_\odot)$&\multicolumn{2}{c}{Open model}
&\multicolumn{2}{c}{Closed model}\\
   & Stars & Gas & Stars & Gas \\
\hline
10 & -0.4 & -0.3  & 0.2 &  0.1  \\
11 & -0.1 & -0.1  & 0.2 &  0.1  \\
12 &  0.1 &  0.1  & 0.2 &  0.1  \\
\hline
\end{tabular}
\end{table}

In Fig.~\ref{imlr} we show iron and silicon mass in the clusters versus
partial luminosity of spirals, ellipticals, and lenticulars (all data
needed to compute these quantities were taken from Fukazawa et al. and
Arnaud et al.). These two plots reveal the surprising fact: both
iron and silicon masses follow nearly the same path for all morphological
types, with the only exception of \object{A426} (Perseus cluster). It is
noteworthy that according to Andreon (\cite{A94}) the number of spirals
in this cluster is underestimated by a factor of 2 or 3.

\begin{figure}
\resizebox{\hsize}{!}{\includegraphics{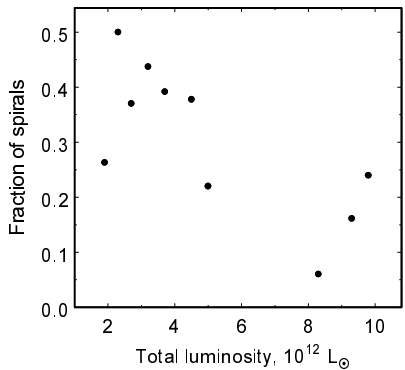}}
\caption{Fraction of spirals as a function of total cluster luminosity.}
\label{fracspil}
\end{figure}

However, one might argue that the fact that iron (or silicon) mass in
the cluster correlates with a partial luminosity of any morphological
type is not so meaningful. Renzini et al. (\cite{R93}) suggested that
it is the ratio of iron mass to cluster luminosity (IMLR) that does
make sense in studying the ICM chemical evolution. If we plot fraction
of spirals versus total luminosity for galaxies from Arnaud et al.
(\cite{Aetal92}) sample (Fig.~\ref{fracspil}), we see that in the
range of luminosities from $10^{12}$ to $10^{13}L_\odot$ fraction of
spirals drops from 0.5 to approximately 0.1. One might expect that the
IMLR should decrease in spiral-rich clusters because of decreasing
number of metal sources.

\begin{figure*}
\resizebox{0.8\hsize}{!}{\includegraphics{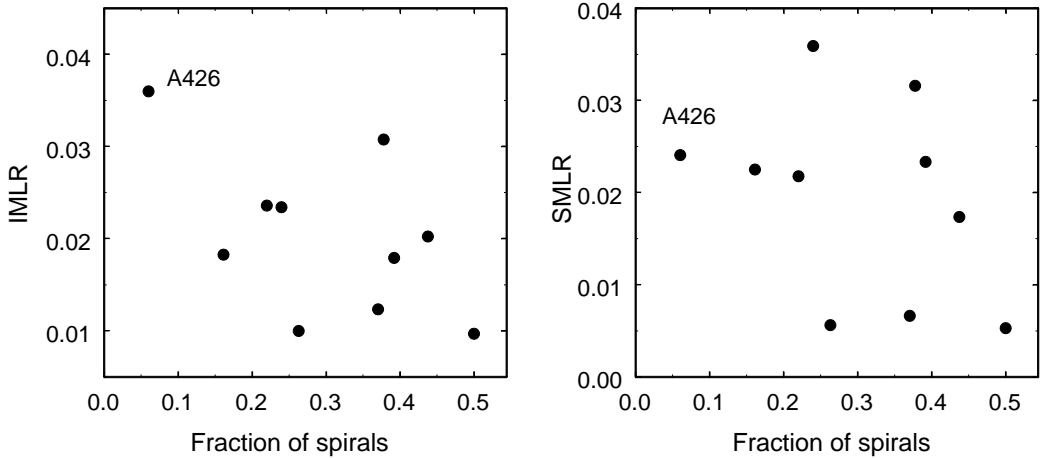}}
\caption{Iron mass-to-light ratio (a) and silicon mass-to-light ratio (b) for
a sample of galaxy clusters.}
\label{frac}
\end{figure*}

To check if this is the case, we plot in Fig.~\ref{frac} the ICM iron
mass to light ratio (IMLR) and silicon mass to light ratio (SMLR)
versus fraction of spirals in galaxy clusters. In both cases there is
an apparent correlation between these two values. However, it rests
mainly upon the cluster \object{A426} which we have already mentioned.
So, if we drop this cluster from consideration, Figs~\ref{imlr} and
\ref{frac} reveal that, first, iron and silicon masses in the galaxy
clusters increase with the luminosity of \emph{any} morphological
type. Second, there is no clear correlation between the IMLR (or SMLR)
and a fraction of spirals in the cluster.

\section{Conclusion}

In this paper we investigate the role played by disk galaxies in the
chemical evolution of the ICM. We show that if a mass of the galaxy
does not exceed $\sim5\cdot10^{11} M_\odot$ it contributes nearly the
same amount of metals into the ICM as an elliptical galaxy of
comparable mass. However, we have found that massive spirals are able
to retain their heavy elements and thus do not participate in the ICM
enrichment. Integration of the heavy element yields of galaxies over
the galactic mass spectrum shows that while ellipticals do play the
determining role in the chemical evolution of rich clusters, spirals
can be significant source of $Z$ in poor clusters.

We should mention briefly some moments which are important for future
improvement of our model: there are evidences that the galactic mass
function changed in time (e.g. because of coalescense of early
galaxies); new important data on abundance of other chemical elements
are expected from space experiments; dynamical interaction of gas and
galaxies in the clusters seems to be important and it should be
investigated in more details.

\begin{acknowledgements}
This work was supported by the Russian Foundation of Basic Researches
grant number 96-02-16351.
\end{acknowledgements}

\end{document}